\documentclass{appolb}
\usepackage{epsfig,picins}

\begin{document}
\title{Production of Protons in Photoproduction at HERA
\thanks{Presented at the X International Workshop on Deep Inelastic Scattering
(DIS2002) in Cracow, Poland, 30 April - 4 May 2002}
}

\author{Dmitry Ozerov
\address{Institute for Theoretical and Experimental Physics, ITEP, \\
B.Cheremushkinskaya 25, Moscow 117259, Russia \\
\vspace{0.3cm}
On behalf on the H1 Collaboration
}
}

\maketitle
\begin{abstract}
 A measurement of the inclusive cross-section for the photoproduction
of protons in the central fragmentation region at HERA is presented. The measured
cross-section is compared with the prediction of the PYTHIA MC model.
The obtained value of the diquark suppression factor in the Lund string model
is different from that obtained in $e^+e^-$ data.
The dependence of the cross-section as a
function of the transverse mass $M_T$  is shown to be similar
for different hadrons produced in photoproduction at HERA.
\end{abstract}

  
\section{Introduction}

 Production of low momentum particles in the region away
from the beam remnant is a key laboratory to study the non-perturbative
effect of hadronisation, \ie the process of conversion of the partonic
final state into hadrons. The lack of understanding of this phenomenon
leads to a huge variety of phenomenological models, each with
a large number of free parameters. An example of such a model
is the Lund string model \cite{lund} and its implementation in
the JETSET Monte Carlo (MC) model\cite{pythia}. 
Comparison of the predictions of this
model with accurate data from the LEP experiments \cite{delphi}
allows some of the parameters to be precisely fixed.
It is possible to test the universality of the fragmentation
prescription in such models, by comparing the predictions of the MC,
tuned with $e^+e^-$ data, to the HERA data in photoproduction. In this paper
we present a measurement\cite{h1:proton} of centrally produced low momentum
protons and compare the results with the predictions of 
the PYTHIA MC \cite{pythia}. A common behaviour as a function of
$M_T =m+p_T$ in the production cross-section
for different hadrons of mass $m$ and transverse momentum $p_T$, 
produced in photoproduction, is also demonstrated.

\section{Event Selection}

 Experimental data for the analysis were collected with the H1 detector
\cite{h1detector} during the 1996 running period, in which HERA collided
27.5 GeV positrons with 820 GeV protons. The data sample corresponds 
to an integrated luminosity of 6.0 pb$^{-1}$. 
 
 The photoproduction events are identified by a first
level trigger which requires energy in the small angle electron tagger and
the presence in the central tracking chamber of several well measured tracks
which originate from the primary vertex. The former requirement restricts the
kinematic region to $Q^2 < 0.01$ GeV$^2$ and $0.3 < y_e < 0.7$, where $Q^2$ 
is the virtuality of the exchanged photon and $y_e$ is the inelasticity of
the scattered positron. The average centre of mass energy of the 
$\gamma p$ system is 200 GeV. 

 Background events coming from the interaction of beam particles with 
the residual gas in the beam pipe (beam-gas events) are characterized
by small or zero energy deposition in the direction opposite to the
direction of the incoming beam particle. 
To suppress such background, only events with energy 
deposits greater than 2 GeV in both the forward region of the 
Liquid Argon Calorimeter and in the backward calorimeter SPACAL
are accepted.

\piccaption[]{DCA distribution for a) identified anti-protons and 
b) identified protons.
  \label{fig:dca}
}
\parpic(6.68cm,5.5cm)[r]{
  \begin{picture}(6.68,5.5)(0.0,0.1)
    \epsfig{file=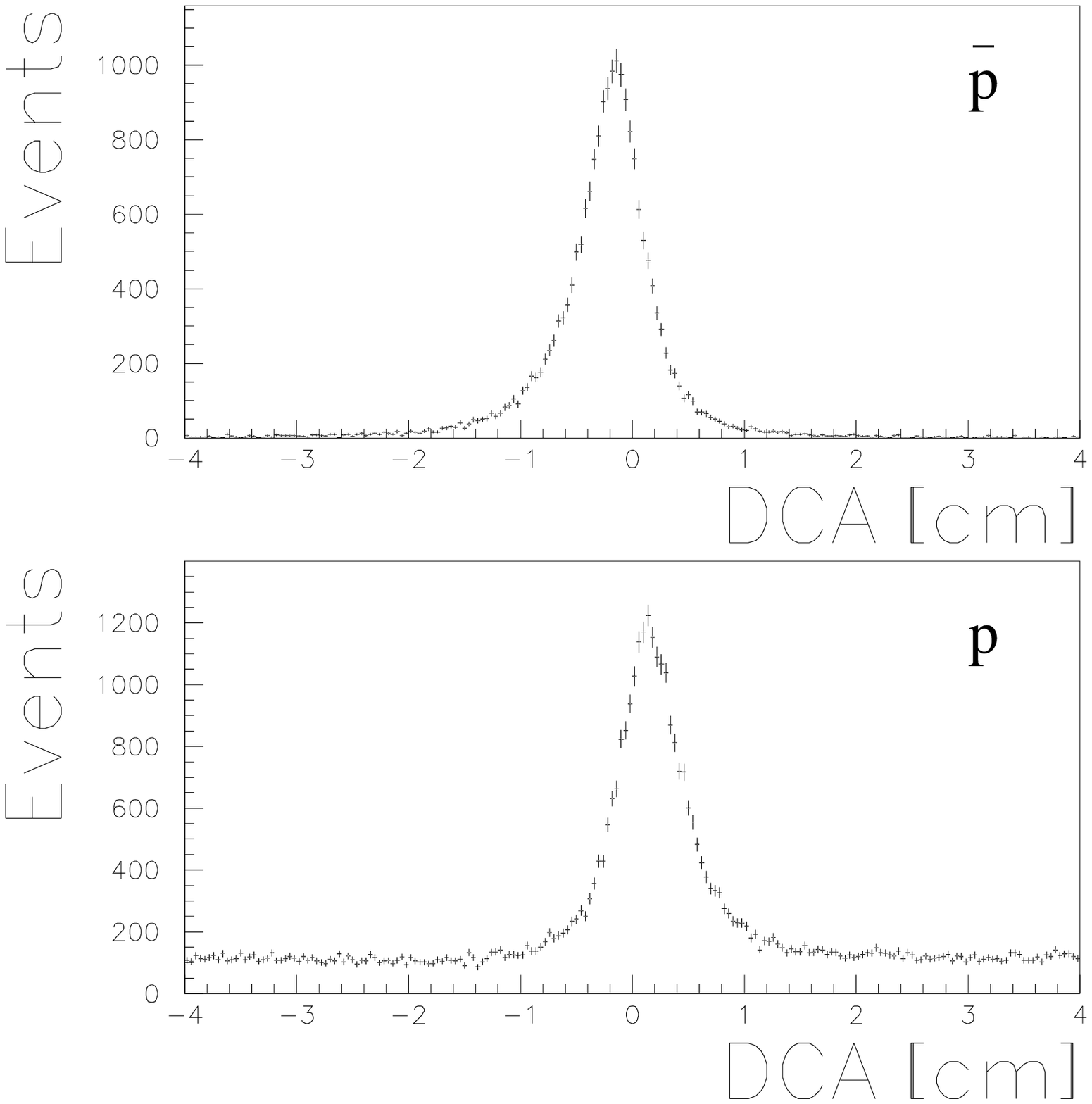,width=6.5cm,height=6.3cm}
    \put(-5.0,4.7){a)}
    \put(-5.0,2.0){b)}
  \end{picture}
}

\noindent Charged particles are measured in the 
H1 central tracker which covers the
polar angle range of $22^0 < \theta < 150^0$, where $\theta$ is defined
with respect to the incoming proton direction. $dE/dx$, the measured specific 
ionization energy loss of the particle,
is used to identify the protons
in the region of high acceptance and resolution, $0.3 < p_T < 0.55$ GeV and
$-0.3 < y < 0.3$, where $p_T$ is the transverse momentum relative to the
beam-line and $y$ is the rapidity of the particle. 

The main source of the background is the production of
secondary protons by particles from ep collision, interacting
in the beam pipe and other material in front of the track chambers.
In order to estimate the
yield of such protons the distribution of closest approach (DCA) of the
track to the primary vertex is analyzed. For background protons this 
distribution is flat, in contrast to the shape for protons and 
anti-protons from ep interaction, see Fig.\ref{fig:dca}. This
background is subtracted on a statistical basis.
 
\section{Results}

\begin{figure}
\hspace*{-0.5cm}
\epsfig{file=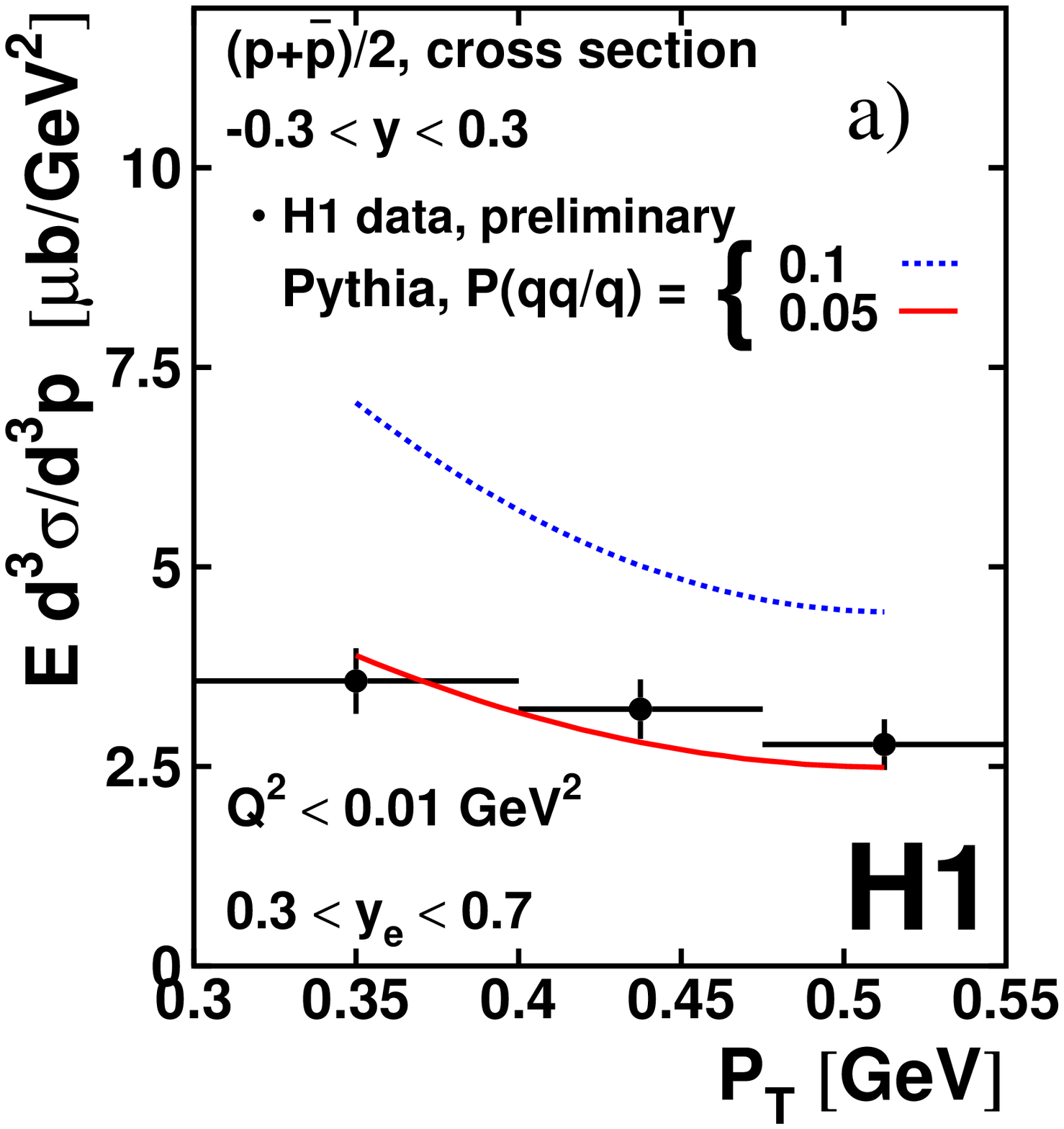,width=0.54\textwidth,height=0.45\textheight}
\epsfig{file=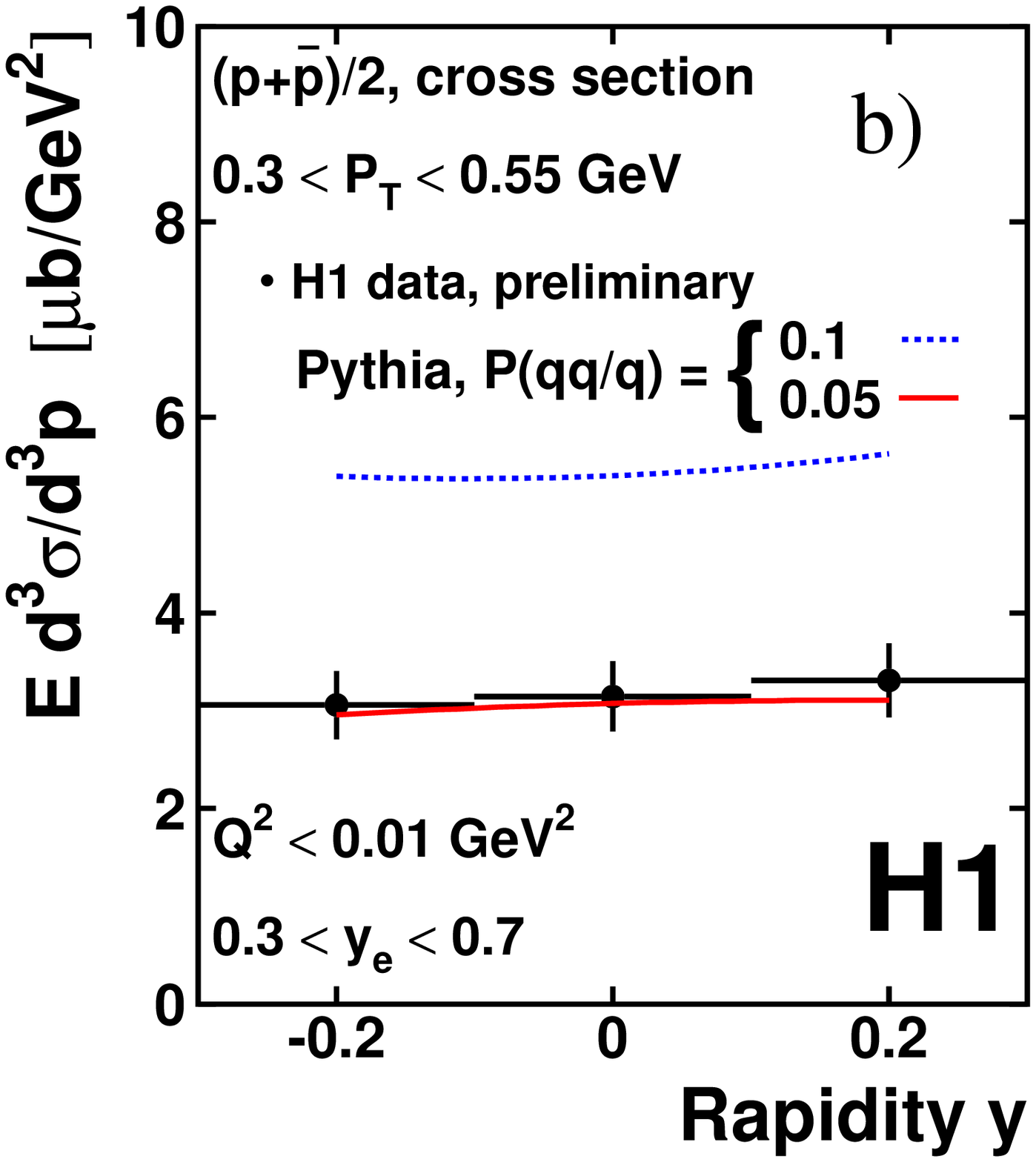,width=0.54\textwidth,height=0.45\textheight}

\caption{The measured photoproduction cross-section 
$E\frac{d^3\sigma}{d^3p}$ as a function
of a) transverse momentum $p_T$ and b) rapidity $y$, in the region of 
$0.3 < p_T < 0.55$ GeV and $-0.3 < y < 0.3$. The cross-section represents
half of the sum of protons and anti-protons inclusively produced.
The prediction of the PYTHIA model is shown
with two different values of the diquark suppression factor, 0.1 (upper
curve) and 0.05 (lower curve).}
\label{fig:data_pythia}
\end{figure}

 The measurement is made in the laboratory frame of reference and is 
presented
as an average cross-section $E\frac{d^3\sigma}{d^3p}$ for photoproduction
of protons and anti-protons, \ie as half the sum of the corresponding 
proton and  anti-proton cross-sections. 
Here E and p are the proton\footnote{Henceforth
proton is used to designate both proton and anti-proton.} energy and
momentum, respectively.
This cross section is shown as a function of
the proton transverse momentum $p_T$ and rapidity $y$ in 
Fig.\ref{fig:data_pythia}. The errors are the quadratic sum of the 
statistical and systematic uncertainties. The systematic component is
dominant. Using the MC simulation, the resulting cross-sections are
corrected for the acceptance cuts discussed in the previous section (cuts
on the energy in forward and backward part of the H1 calorimeters and 
on the number of tracks coming from the primary vertex).

 The measured cross-section is compared with the PYTHIA MC model\cite{pythia}
prediction, including the GRV-LO proton and photon parton 
densities \cite{grv}. In this model leading order (LO) QCD matrix elements 
are used to describe the hard scattering processes, while parton shower
algorithms produce initial and final state parton radiation. Multiple
soft and hard parton interactions are also simulated. For the hadronisation
step the Lund string model \cite{lund}, as implemented in 
JETSET 7.4\cite{pythia}, was utilized. 

 The yield of baryons produced during the hadronisation in the Lund
string model is proportional
to the amount of diquark states created in the color field. Thus the most
important parameter of this model related to the production of
protons is the diquark suppression factor which determines the probability
of producing a ($qq,\bar{q}\bar{q}$) pair relative to the ($q,\bar{q}$) 
pair in the color field. 
Results from $e^+e^-$ collisions at LEP favor a value of 
approximately 0.1\cite{delphi} for this parameter.
 
 The PYTHIA MC model with the $e^+e^-$-derived value of 0.1
overshoots the data (Fig.\ref{fig:data_pythia}), although it 
provides a fair
description of the shape of the spectra. With the parameter value 0.05 
both the 
shape and yield are well described. This could indicate the 
absence of universality of baryon production within this model approach.
More detailed studies of the production of different hadrons over
a wider range a phase space than covered in this work are required
in order to further understand this phenomenon.

\section{Other identified particles}

 The present measurement of the inclusive proton cross-section adds a
new point to the 
series of measurements on inclusive hadron production
in photoproduction at HERA. Previous data on the production of
long-lived low mass hadrons 
$\pi^\pm$\cite{h1:pions,zeus:pions}, $K^0$, $\Lambda^0$ \cite{h1:k0_lambda}
and the charmed mesons $D^{*+}$ and $D_s$\cite{zeus:charm,h1:charm} 
are available.
In the following we compare these cross-sections with the present
new measurement of protons. For this comparison we
convert the published data points to differential cross-sections
$E\frac{d^3\sigma}{d^3p}$, correcting for the flux of the photons\cite{ww:flux}
where ep cross-sections are given.
The pion data are recalculated from the measured charged particle
spectra taking into account a 17$\%$ admixture of charged kaons
and protons to these spectra. In the comparison for different
hadrons the cross-sections are given for one isospin projection and
are corrected for the particle spin.

\begin{figure}
\hspace*{-0.5cm}
\epsfig{file=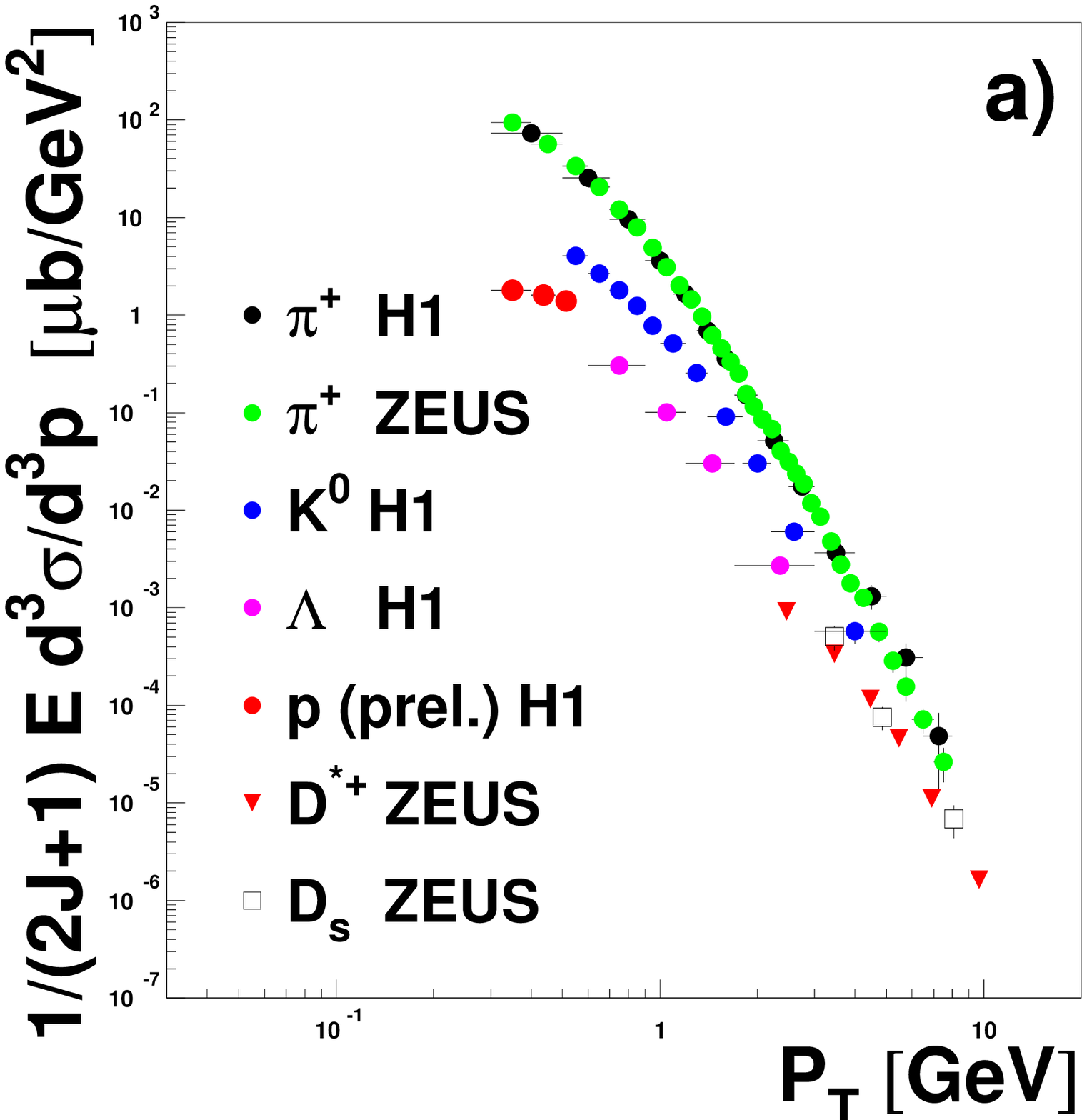,width=0.54\textwidth,height=0.43\textheight}
\epsfig{file=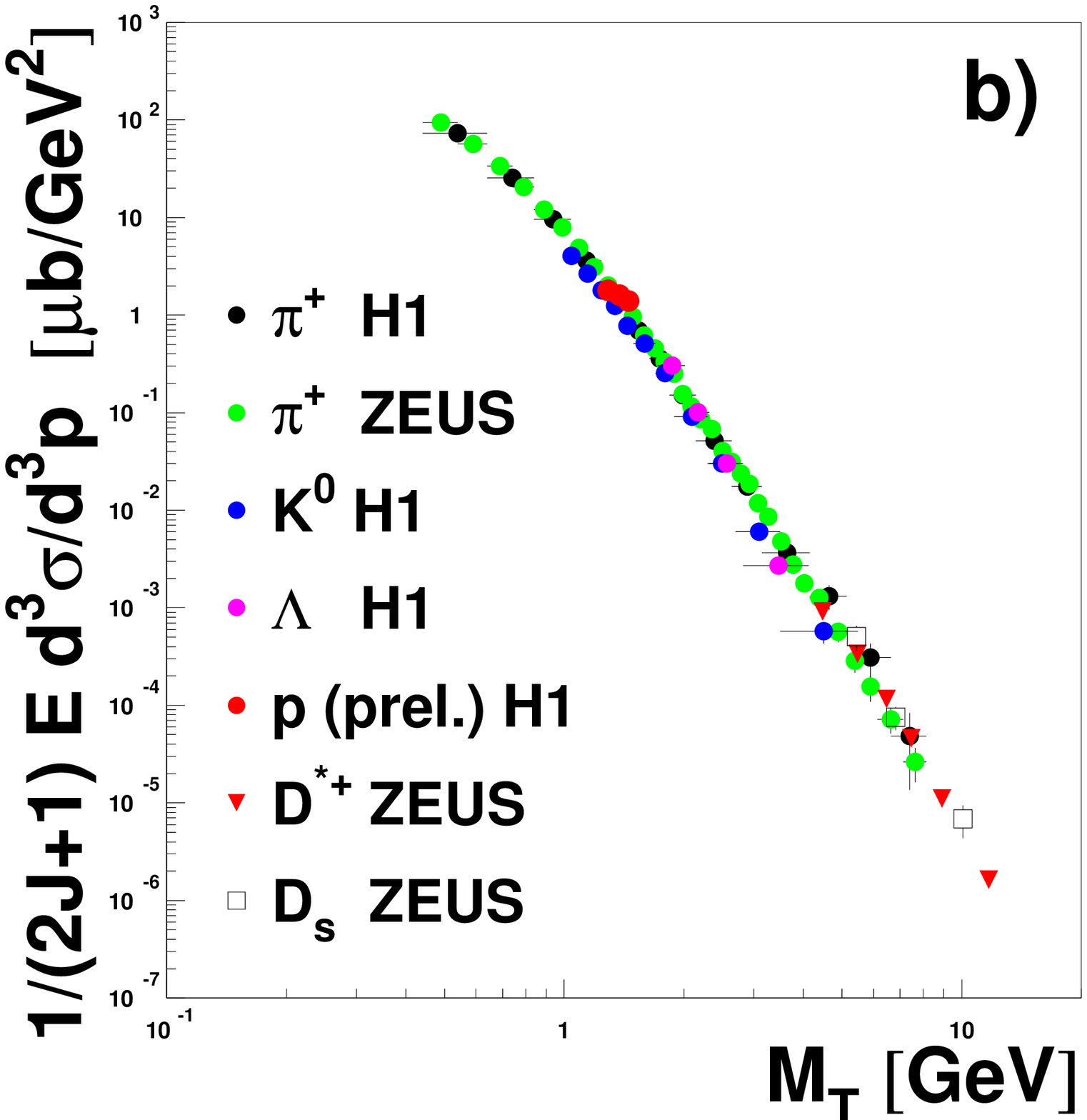,width=0.54\textwidth,height=0.43\textheight}

\caption{Photoproduction cross section $E\frac{d^3\sigma}{d^3p}$ 
for different hadrons as a function of a) transverse momentum
$p_T$ and b) transverse mass $M_T$ ($=m+p_T$), in the central
rapidity region as measured in the laboratory frame of reference. The
cross-sections are given for one isospin and corrected for the spin.}
\label{fig:pt_mt_dependence}
\end{figure}

In Fig.\ref{fig:pt_mt_dependence}a these
cross-sections are plotted as a function of the transverse momentum $p_T$.
The individual particle cross-section clearly
depends strongly on the specific hadron mass. 
It was shown in lepton-hadron\cite{h1:pions} and 
hadron-hadron\cite{ua1:hadrons} collisions that
the $p_T$-dependence of the hadron cross-sections is well described by a
power-law behavior $A(p_0+p_T)^{-n}$. In $e^+e^-$ collisions, on the 
other hand, the production rates of hadrons are determined only
by the particle spins, isospins and masses\cite{lep:uvarov}. 
We replot in Fig.\ref{fig:pt_mt_dependence}b our cross-section data 
as a function of the modified 
transverse mass $M_T$ ($=m+p_T$, where $m$ is the particle mass).
The data show that the cross-sections for different
hadrons produced in the photoproduction regime at HERA have
similar behaviour as a function of this variable. 

 In summary, we have presented a measurement of the production of protons
in photoproduction at HERA, performed in the central rapidity region
and at low proton momentum. The PYTHIA MC model only describes the 
normalisation of the cross section if the model parameter 
(diquark suppression factor) is
a factor of 2 smaller than
the value obtained from $e^+e^-$ collisions.
The cross-sections for different hadrons produced in photoproduction 
at 200 GeV are shown to have similar behaviour as a function of
$m+p_T$.

I would like to thank Grazyna Nowak and 
the whole local organizing committee for the kind hospitality and
for the high level of organization of this very fruitful workshop.
I wish to acknowledge Jan Olsson and Paul Richard Newman for the 
careful reading of the manuscript and for useful discussions.

\end{document}